# On Universal Vassiliev Invariants

Daniel Altschuler [*]

*Institut für Theoretische Physik*

*ETH-Hönggerberg*

*CH-8093 Zürich, Switzerland*

and

Laurent Freidel

*Laboratoire de Physique Théorique ENSLAPP* [†]

*Ecole Normale Supérieure de Lyon*

*46, allée d'Italie, 69364 Lyon Cedex 07, France*

## Abstract

Using properties of ordered exponentials and the definition of the Drinfeld associator as a monodromy operator for the Knizhnik-Zamolodchikov equations, we prove that the analytic and the combinatorial definitions of the universal Vassiliev invariants of links are equivalent.

hep-th/9403053  8 Mar 94

[*]Supported by Fonds national suisse de la recherche scientifique.

[†]URA 1436 du CNRS, associée à l'Ecole Normale Supérieure de Lyon et au laboratoire d'Annecy-le-Vieux de Physique des Particules.

# 1 Introduction

Vassiliev's knot invariants [1] contain all the invariants, such as the Jones [2], HOMFLY [3] and Kauffman [4] polynomials, which can be obtained from a deformation $U_h(\mathcal{G})$, usually called quantum group [5], of the Hopf algebra structure of enveloping algebras $U(\mathcal{G})$, where $\mathcal{G}$ is a semisimple Lie algebra.

For a compact semisimple Lie group $G$ with Lie algebra $\mathcal{G}$, observables of the quantized Chern-Simons model give knot invariants [6] associated to $U_h(\mathcal{G})$ at special values $h = 2\pi i\, k^{-1}$, $k$ a positive integer. The coefficients of the expansion in powers of $h$ of these observables are examples of Vassiliev invariants. This is a particular case of a general theorem [7], which states that for all $h$ the coefficients of the power series expansion of the invariants associated with semisimple Lie algebras are Vassiliev invariants.

By treating the Chern-Simons model with the conventional methods of perturbation theory, the coefficients of the powers of $h$ of the observables can be computed [8]. Feynman diagrams and Feynman rules are the main tools of the computation. Given a knot, or more generally a link $L$, and the degree $n$ (order in perturbation theory) or power of $h$ in which one is interested, the corresponding invariant $V_n(L)$ results from the application of a Feynman rule $W_\mathcal{G}$ to a finite linear combination $Z_n^{CS}(L)$ of diagrams. The vector space $D_n$ of diagrams of degree $n$ is of finite dimension, and $Z_n^{CS}(L) \in D_n$ depends on $L$ and on the form of the Chern-Simons action. The Feynman rule $W_\mathcal{G}$ depends on $\mathcal{G}$ and the representations occuring in the definition of the observables. It is an element of the dual space $D_n^*$, and $V_n(L) = \langle W, Z_n^{CS}(L) \rangle$.

Here we have used Bar-Natan's way [9] of describing Feynman rules and diagrams. He found that the diagrams and rules of Chern-Simons theory obey a small number of fundamental properties, and this led him to define general diagrams and rules, the latter which he called weight systems, by these same properties. Kontsevich [10] discovered an integral formula for an invariant $Z_n(L) \in D_n$, which plays for generic $h$ the same role as $Z_n^{CS}(L)$ does for the special values in the Chern-Simons case. The main ingredient hiding behind it is the flat connection associated with the Knizhnik-Zamolodchikov equations. The formal power series $Z(L) = \sum_{n \geq 0} Z_n(L)\, h^n$ is called the universal Vassiliev invariant, since by varying $W$ in $\langle W, Z(L) \rangle$ one gets all the invariants constructed from a deformation of the identity solution of the Yang-Baxter equation.

The deep questions remain: do the Vassiliev invariants form a complete set of knot invariants? Are there any other Feynman rules (weight systems) than those of the type $W_\mathcal{G}$ associated to Lie algebras? The following troubling result is related to the second question: Vassiliev invariants are invariants of oriented knots. However all Vassiliev invariants $\langle W_{\mathcal{G}}, Z_n(L) \rangle$ are independent of the orientation. Is there any weight system $W$ which can distinguish the two orientations of a knot? The simplest example of a knot which is not isotopy-equivalent to the same knot with the reversed orientation was found in [11]. It has 15 crossings.

The knot invariants constructed using the representations of $U_h(\mathcal{G})$ have been generalized to all quasi-triangular Hopf algebras by Reshetikhin and Turaev [12]. Their construction is purely combinatorial, the proof of invariance consists in verifying that the Reidemeister moves do not change the relevant expressions. Recently, similar combinatorial definitions of universal Vassiliev invariants have appeared [13, 14]. The aim of this paper is to show that the combinatorial and the analytic definition of Kontsevich are equivalent. More precisely, since the combinatorial approach leads naturally to invariants of framed knots, we will show that it is equivalent to a variant of the Kontsevich formula, which was written originally for unframed knots. The same notion of Kontsevich integral for framed knots appears in [15]. However here we will define it in a way which does not require the framed knot to be presented as a product of tangles with special properties.

As we were finishing this paper, we learned that the equivalence of the combinatorial and analytic definitions had been shown before in [16]. We believe that our methods makes the proof more direct. While the authors of [16] work with the individual terms $Z_n(L)$ which are iterated integrals and are led to long computations in order to identify these terms with the corresponding terms of the combinatorial invariants, we essentially treat the whole series $Z(L)$ at once. It turns out that the main contribution to $Z(L)$ is a type of series called ordered exponential in the physics literature. Ordered exponentials satisfy many interesting, but not well-known, identities which makes them very powerful. A crucial step in the proof of equivalence is to identify an expression for the Drinfeld associator [17, 18] among the Kontsevich integrals. We do it quite naturally using only Drinfeld's definition of the associator as a monodromy operator between solutions of the Knizhnik-Zamolodchikov differential equations. We don't have to find before some expressions for the coefficients of the associator viewed as a power series, as is done in [16].

The contents of the paper are as follows. In section 2, we define the ordered exponential and prove the properties which we use later in the proof. Sections 3 and 4 are devoted to the definitions of the combinatorial invariants and the Bar-Natan (Feynman) diagrams. In section 5 we define the Kontsevich integral of framed links. The proof of the equivalence theorem occupies sections 6 and 7.

## 2  The ordered exponential

We give here the properties of the ordered exponential which we will need later in the paper. Let $A : \mathbb{R} \to \mathcal{A}$ be a function with values in the associative algebra $\mathcal{A}$. The ordered exponential of $A$:

$$g(x,y) = \overleftarrow{\exp} \int_y^x du\, A(u), \tag{2.1}$$

is the solution of the differential equation:

$$\frac{\partial}{\partial x} g(x,y) = A(x) g(x,y) \tag{2.2}$$

with the initial condition $g(y,y) = 1$. An equivalent definition is

$$g(x,y) = 1 + \sum_{n=1}^{+\infty} \int_y^x dt_1 \int_y^{t_1} dt_2 \ldots \int_y^{t_{n-1}} dt_n\, A(t_1) A(t_2) \cdots A(t_n). \tag{2.3}$$

**Proposition 1** *The ordered exponential is multiplicative:*

$$g(x,y) = g(x,z) g(z,y). \tag{2.4}$$

*Proof.* Let $\tilde{g}(x,y)$ be the solution of the equation:

$$\frac{\partial}{\partial x} \tilde{g}(x,y) = -\tilde{g}(x,y) A(x) \tag{2.5}$$

with the initial condition $\tilde{g}(y,y) = 1$. Then $\tilde{g}(x,y)$ is a left inverse of $g(x,y)$. Indeed we have $\partial_x (\tilde{g}(x,y) g(x,y)) = 0$, thus $\tilde{g}(x,y) g(x,y) = 1$. The functions $g$ and $\tilde{g}$ also satisfy the relation:

$$g(x,y) \tilde{g}(x,z) = g(x,z) g(z,y) \tilde{g}(x,z), \tag{2.6}$$

which follows from the fact that the two sides of the equality satisfy the same differential equation with respect to $x$ and the same initial condition at $x = z$. Multiplying on the right by $g(x,z)$ we obtain the desired result. □

**Corollary 1** $g(x,y)$ *is invertible,* $g^{-1}(x,y) = g(y,x)$ *and*

$$\frac{\partial}{\partial y} g(x,y) = -g(x,y) A(y). \tag{2.7}$$

**Proposition 2** *Let* $\delta \in \mathrm{der}\mathcal{A}$ *be a derivation of* $\mathcal{A}$, *then:*

$$\delta g(x,y) = \int_y^x dt\, g(x,t) \delta A(t) g(t,y). \tag{2.8}$$

**Proposition 3** *Behaviour with respect to gauge transformations: if $h : \mathbb{R} \to \mathcal{A}$ is a function such that $h(t)$ is invertible for $x \geq t \geq y$, then*

$$h(x) \left( \overleftarrow{\exp} \int_y^x dt A(t) \right) h^{-1}(y) = \overleftarrow{\exp} \int_y^x dt \left( h(t) A(t) h^{-1}(t) + \partial_t h \, h^{-1}(t) \right). \tag{2.9}$$

**Proposition 4** *Factorization identities:*

$$\overleftarrow{\exp} \int_y^x dt (A(t) + B(t)) = \overleftarrow{\exp}(\int_y^x dt A(t)) \overleftarrow{\exp}(\int_y^x dt \, {}^A B(t)) \tag{2.10}$$

$$\overleftarrow{\exp} \int_y^x dt (A(t) + B(t)) = \overleftarrow{\exp}(\int_y^x dt B^A(t)) \overleftarrow{\exp}(\int_y^x dt A(t)) \tag{2.11}$$

$$\left( \overleftarrow{\exp} \int_t^x du A(u) \right) B(t) \left( \overleftarrow{\exp} \int_t^x du A(u) \right)^{-1} = \overleftarrow{\exp}(\int_t^x du \, \mathrm{ad} A(u)) \cdot B(t) \tag{2.12}$$

*where*

$$^A B(t) = \left( \overleftarrow{\exp} \int_y^t du A(u) \right)^{-1} B(t) \left( \overleftarrow{\exp} \int_y^t du A(u) \right) \tag{2.13}$$

$$B^A(t) = \left( \overleftarrow{\exp} \int_t^x du A(u) \right) B(t) \left( \overleftarrow{\exp} \int_t^x du A(u) \right)^{-1} \tag{2.14}$$

*and* $\mathrm{ad} : \mathcal{A} \to \mathrm{der} \mathcal{A}$ *is given by* $\mathrm{ad}(a) \cdot b = [a, b] = ab - ba$ *for* $a, b \in \mathcal{A}$.

The proof of propositions 2, 3 and the last identity of proposition 4 consists in checking that the two sides of the equalities satisfy the same differential equation and initial condition. The first two identities of proposition 4 are consequences of proposition 3.

## 3 Ribbon categories

### 3.1 Basic definitions

In this section we recall the definition of the combinatorial invariants. We formulate it in the language of categories and we follow closely the notations of Cartier [14]. Let $\mathcal{C}$ be a monoidal category, and denote the product, which is a bifunctor $\mathcal{C} \times \mathcal{C} \to \mathcal{C}$ by $\otimes$. For any triple of objects $X, Y, Z$ of $\mathcal{C}$, there is an isomorphism $\phi_{X,Y,Z} : (X \otimes Y) \otimes Z \to X \otimes (Y \otimes Z)$, which is natural:

$$(f \otimes (g \otimes h)) \phi_{X,Y,Z} = \phi_{X',Y',Z'} ((f \otimes g) \otimes h) \tag{3.1}$$

for all morphisms $f : X \to X'$, $g : Y \to Y'$, $h : Z \to Z'$, and obeys the pentagon relation:

$$(\mathrm{id}_X \otimes \phi_{Y,Z,T}) \phi_{X,Y \otimes Z,T} (\phi_{X,Y,Z} \otimes \mathrm{id}_T) = \phi_{X,Y,Z \otimes T} \phi_{X \otimes Y,Z,T}. \tag{3.2}$$

The unit object is denoted by $I$, and we will assume that $X \otimes I = I \otimes X = X$ for simplicity.

Let $X_1, X_2, \ldots, X_k$ be a (possibly empty) sequence of objects in $\mathcal{C}$, and consider $X = (((X_1((\otimes(X_2 \otimes \cdots)\cdots)) \otimes X_k))$, with a given distribution of parentheses. We shall say that $X$ is standard if $X = I$, or if all its left parentheses are placed on the left of the first factor $X_1$. We also say that a morphism from $X$ to $X'$ is standard if both $X$ and $X'$ are standard. Every morphism from $I$ to $I$ can be written as the composition of standard morphisms. By Mac Lane's coherence theorem [19], for every object $X$ there is a unique isomorphism $\psi_X$ from $X$ to a standard object $X_{\mathrm{st}}$, and for every pair of morphisms $f: X \to X'$, $g: Y \to Y'$, the morphism $f \otimes_{\mathrm{st}} g = \psi_{X' \otimes Y'}(f \otimes g)\psi_{X \otimes Y}^{-1}$ is standard.

A braiding in a monoidal category $\mathcal{C}$ is a function, which to any pair of objects $X, Y$ of $\mathcal{C}$ associates a natural isomorphism $R_{X,Y}: X \otimes Y \to Y \otimes X$:

$$R_{X',Y'}(f \otimes g) = (g \otimes f) R_{X,Y} \tag{3.3}$$

for all morphisms $f: X \to X'$, $g: Y \to Y'$, satisfying the two hexagon relations

$$R_{X \otimes Y, Z} = \phi_{Z,X,Y}(R_{X,Z} \otimes \mathrm{id}_Y)\phi_{X,Z,Y}^{-1}(\mathrm{id}_X \otimes R_{Y,Z})\phi_{X,Y,Z} \tag{3.4}$$

$$R_{X, Y \otimes Z} = \phi_{Y,Z,X}^{-1}(\mathrm{id}_Y \otimes R_{X,Z})\phi_{Y,X,Z}(R_{X,Y} \otimes \mathrm{id}_Z)\phi_{X,Y,Z}^{-1}. \tag{3.5}$$

Let $X, Y$ be two objects of a monoidal category $\mathcal{C}$. We say that $X$ is a left dual of $Y$ if there are morphisms $a: X \otimes Y \to I$, $b: I \to Y \otimes X$ satisfying the relations

$$(a \otimes_{\mathrm{st}} \mathrm{id}_X)(\mathrm{id}_X \otimes_{\mathrm{st}} b) = \mathrm{id}_X, \tag{3.6}$$

$$(\mathrm{id}_Y \otimes_{\mathrm{st}} a)(b \otimes_{\mathrm{st}} \mathrm{id}_Y) = \mathrm{id}_Y. \tag{3.7}$$

Two left duals of $Y$ are isomorphic. If $a_0: X \otimes Y \to I$, $b_0: I \to Y \otimes X$ are morphisms satisfying the weaker relations

$$(a_0 \otimes_{\mathrm{st}} \mathrm{id}_X)(\mathrm{id}_X \otimes_{\mathrm{st}} b_0) = \lambda_X, \tag{3.8}$$

$$(\mathrm{id}_Y \otimes_{\mathrm{st}} a_0)(b_0 \otimes_{\mathrm{st}} \mathrm{id}_Y) = \lambda_Y, \tag{3.9}$$

where $\lambda_X$ and $\lambda_Y$ are invertible, then $a = a_0(\lambda_X^{-1} \otimes \mathrm{id}_Y)$ and $b = b_0$ satisfy (3.6) and (3.7).

A ribbon category is a braided monoidal category in which each object has a left dual, such that for any object $X$ there is a natural isomorphism $v_X: X \to X$, $f v_X = v_{X'} f$ for all morphisms $f: X \to X'$, satisfying the relations

$$v_{X \otimes Y} = (R_{Y,X} R_{X,Y})^{-1}(v_X \otimes v_Y), \tag{3.10}$$

$$a(v_A \otimes \mathrm{id}_B) = a(\mathrm{id}_A \otimes v_B), \tag{3.11}$$

for all objects $X, Y$ and $A, B$, where $A$ is a left dual of $B$.

## 3.2 Ribbon graphs

An example of ribbon category is $\mathcal{R}$, the category of ribbon graphs [12, 20]. A ribbon graph is a finite set of oriented ribbons in $\mathbb{R}^2 \times [0,1]$, i.e. two-dimensional oriented manifolds with boundaries which are the images of non self-intersecting smooth embeddings $[0,1] \times [0,1] \to \mathbb{R}^2 \times [0,1]$ (open ribbons) or $S^1 \times [0,1] \to \mathbb{R}^2 \times [0,1]$ (annuli). Note that Moebius strips are excluded by this definition so that ribbons have a "white" and a "black" side. We assume that the white side is always facing the observer at the top and bottom of the graph, and that the extremities of all the open ribbons are vertical and lie in $\mathbb{R} \times \{0\} \times \{0,1\}$. Ribbons are also directed, i.e. equipped with an arrow. An example of ribbon graph is shown on figure 1.

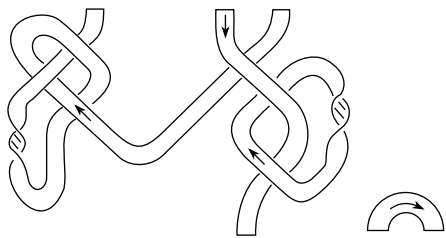

Figure 1:

To each open ribbon we can associate two pairs $\langle x, \alpha \rangle$ and $\langle x', \alpha' \rangle$, called the symbols of the ribbon, where $x, x' \in \mathbb{R}$ and $\alpha, \alpha' \in \{\uparrow, \downarrow\}$. Here $x$ and $x'$ are the middle points of the intervals $[p, q]$ and $[p', q']$ which together form the intersection of the ribbon with $\mathbb{R} \times \{0\} \times \{0,1\}$. The arrows $\alpha$ and $\alpha'$ are given by the direction of the ribbon.

Two graphs are considered equivalent if and only if they are projections of isotopic ribbons. Here by isotopy we mean a smooth isotopy of $\mathbb{R}^3$ which preserves the directions of arrows, the orientation of the graph surface, and keeps the ends of open ribbons fixed. For convenience we will represent pictorially such a ribbon graph as the projection of an oriented link (with possibly open components). This means that we identify the graphs as in figure 2.

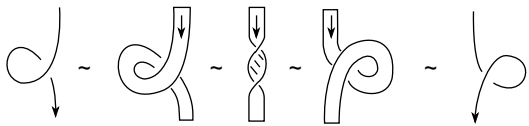

Figure 2:

The objects of $\mathcal{R}$ are words $w$ (formal non-associative expressions) of the form

$$((((e_1 \square ((e_2 \square \cdots )) \cdots ) \square e_k))) \tag{3.12}$$

where each letter $e_i$, $i = 1, \ldots, k$ is a pair $\langle x_i, \alpha_i \rangle$ with $x_i \in \mathbb{R}$ and $\alpha_i \in \{\uparrow, \downarrow\}$. The product of two words $w$, $w'$ is $w \square w'$. The unit object is the empty word $\emptyset$.

A morphism $w \to w'$ in $\mathcal{R}$ is a ribbon graph equipped with an assignment of two words $w$ and $w'$, such that the letters of $w$ ($w'$) are the symbols of the extremities of open ribbons at the bottom (top). Thus $w$ or $w' = \emptyset$ if the graph has no extremities of open ribbons at the bottom or the top. A graph with no open ribbons is called a closed graph. The closed graphs are the morphisms from $\emptyset$ to $\emptyset$. Note that a closed graph defines a framed link, and vice-versa. The tensor product of two graphs $C : w \to w'$, $C' : x \to x'$, denoted $C \square C' : w \square x \to w' \square x'$, is defined by placing them side by side, as in figure 3.

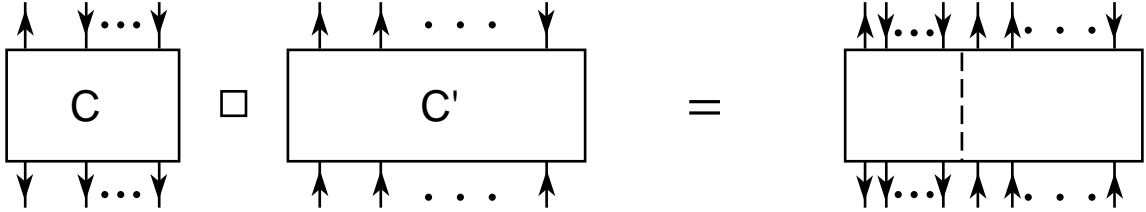

Figure 3:

It will be important for us that every closed graph is a composition of elementary standard graphs $X_{i,n}^\pm$, $\bigcap_{i,n}$ and $\bigcup_{i,n}$ (fig. 4, 5, 6, 7).

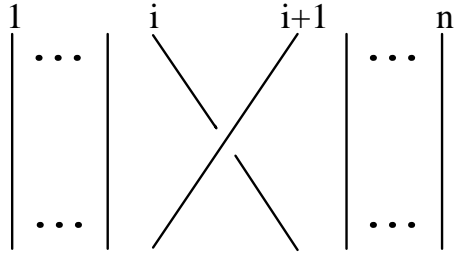

Figure 4: $X_{i,n}^+$

The morphism $\phi_{w,w',w''}$ is shown on figure 8, the braiding $R_{w,w'}$ on figure 9. A fat line stands for a number of thin parallel lines close to each other representing a word. The left dual of the word $w$ is the word $w'$ obtained by reversing the order of the letters and the parentheses, i.e. reading $w$ from right to left, and then reversing the direction of the arrows of each letter. The morphisms $a : w'\square w \to \emptyset$ and $b : \emptyset \to w\square w'$ are shown on figure 10. The relations (3.6) and (3.7) are translated on figure 11. The morphism $v_w$ and its inverse are given on figure 12.

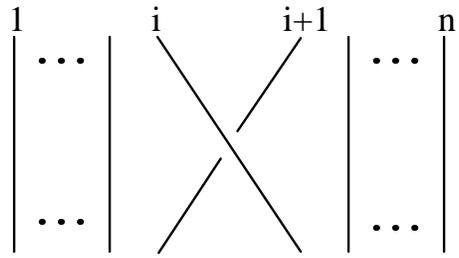

Figure 5: $X_{i,n}^-$

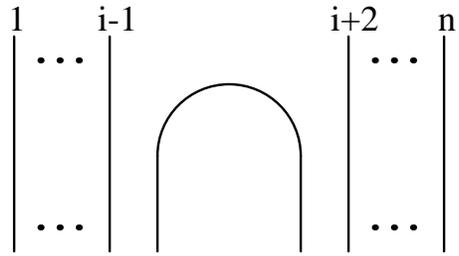

Figure 6: $\bigcap_{i,n}$

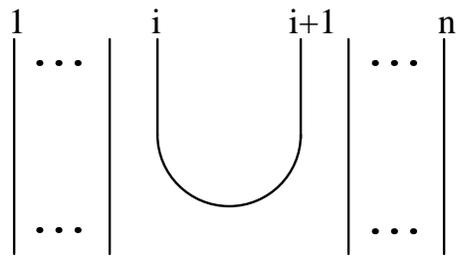

Figure 7: $\bigcup_{i,n}$

$$\Phi_{\omega\,\omega'\,\omega''} = \begin{array}{c} \omega \;\square\; (\omega' \;\square\; \omega'') \\ \Big| \quad \Big| \quad \Big| \\ (\omega \;\square\; \omega') \;\square\; \omega'' \end{array}$$

Figure 8:

$$R_{\omega\omega'} = \begin{array}{c} \omega' \;\square\; \omega \\ \diagdown\!\!\!\diagup \\ \diagup\!\!\!\diagdown \\ \omega \;\square\; \omega' \end{array}$$

Figure 9:

$$a = \bigcap_{\omega' \quad \omega} \qquad b = \bigcup^{\omega \quad \omega'}$$

Figure 10:

## 3.3 Tensor categories

Let $K$ be a field of characteristic 0. A tensor category $\mathcal{T}$ is a strict monoidal category, i.e. $\phi = \text{id}$, such that for all objects $X, Y$ the morphisms $X \to Y$ form a $K$-vector space and the tensor product of morphisms $f \otimes g$ is bilinear, equipped with a symmetric braiding $\sigma$ ($\sigma_{Y,X} = \sigma_{X,Y}^{-1}$), and in which every object has a left dual.

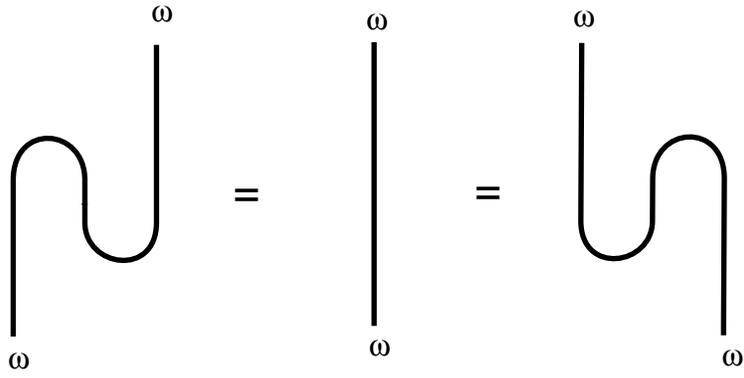

Figure 11:

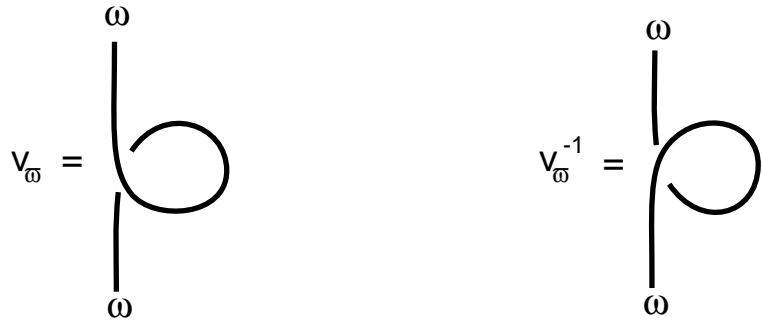

Figure 12:

An infinitesimal braiding $t$ in $\mathcal{T}$ is a function which to any pair $X, Y$ of objects associates a natural morphism $t_{X,Y} : X \otimes Y \to X \otimes Y$ satisfying

$$\sigma_{X,Y}\, t_{X,Y} = t_{Y,X}\, \sigma_{X,Y}, \tag{3.13}$$

$$t_{X \otimes Y, Z} = \mathrm{id}_X \otimes t_{Y,Z} + p^{-1}(t_{X,Z} \otimes \mathrm{id}_Y)p, \tag{3.14}$$

where $p = \mathrm{id}_X \otimes \sigma_{Y,Z}$. For example, if $\mathcal{G}$ is a finite-dimensional semisimple Lie algebra, there is an infinitesimal braiding in the category of finite-dimensional $\mathcal{G}$-modules given by

$$t_{X,Y} = \sum_a (e_a)_X \otimes (e^a)_Y, \tag{3.15}$$

where $\{e_a\}$ is a basis of $\mathcal{G}$, $\{e^a\}$ is the dual basis with respect to the Killing form, and $(e_a)_X$ denotes the action of $e_a$ on $X$.

Let $\mathcal{B}_m$ be the associative graded algebra over $K$ with generators $t_{ij} = t_{ji}$, $i \neq j$, $i, j \in \{1, \ldots, m\}$ of degree 1, and relations

$$[t_{ij}, t_{ik} + t_{jk}] = 0, \tag{3.16}$$

$$[t_{ij}, t_{kl}] = 0, \tag{3.17}$$

where $i, j, k, l$ are distinct. Denote by $\widehat{\mathcal{B}_m}$ the completion with respect to the topology defined by the gradation. Similarly, let $\mathcal{F}_2$ be the free associative graded algebra on two generators $A_1, A_2$ of degree 1, and let $\widehat{\mathcal{F}_2}$ be its completion. A Drinfeld series, or associator, is a formal non-commutative series $\Phi(A_1, A_2) \in \widehat{\mathcal{F}_2}$ satisfying

$$\Phi(A_1, A_2) = \Phi(A_2, A_1)^{-1}, \tag{3.18}$$

$$\Phi(t_{12}, t_{23} + t_{34})\, \Phi(t_{13} + t_{23}, t_{34}) = \Phi(t_{23}, t_{34})\, \Phi(t_{12} + t_{13}, t_{24} + t_{34})\, \Phi(t_{12}, t_{23}), \tag{3.19}$$

$$\exp\frac{1}{2}(t_{13} + t_{23}) = \Phi(t_{13}, t_{12}) \exp\frac{1}{2}(t_{13})\, \Phi(t_{13}, t_{23})^{-1} \exp\frac{1}{2}(t_{23})\, \Phi(t_{12}, t_{23}), \tag{3.20}$$

where the last two relations hold in $\widehat{\mathcal{B}_4}$. Drinfeld has shown [17, 18] that such series exist for all $K$, in particular $K = \mathbb{Q}$. For $K = \mathbb{C}$ he gave an explicit construction of a solution $\Phi_{KZ}$ using the properties of the Knizhnik-Zamolodchikov equations: $\Phi_{KZ}(A_1, A_2) = G_2^{-1} G_1$, where $G_1, G_2 \in \widehat{\mathcal{F}_2}$ are two solutions of the differential equation

$$G'(x) = \frac{1}{2\pi i}\left(\frac{A_1}{x} + \frac{A_2}{x-1}\right) G(x) \tag{3.21}$$

defined in $0 < x < 1$ with the asymptotic behaviour $G_1(x) \sim x^{A_1/2\pi i}$ for $x \to 0$ and $G_2(x) \sim (1-x)^{A_2/2\pi i}$ for $x \to 1$. The coefficients of $\Phi_{KZ}$ are given by generalizations of Riemann's $\zeta$ function [15].

Note that $\widehat{\mathcal{F}}_2$ becomes a topological Hopf algebra with the comultiplication defined by $\Delta(A_i) = A_i \otimes 1 + 1 \otimes A_i$, $i = 1, 2$. Let $\mathcal{L}$ be the Lie algebra of primitive elements in $\widehat{\mathcal{F}}_2$, and let $\mathcal{L}' = [\mathcal{L}, \mathcal{L}]$ be the derived subalgebra. Then $\log \Phi_{KZ}(A_1, A_2) \in \mathcal{L}'$, and if $\Phi(A_1, A_2)$ is a solution of (3.18-3.20) of the form $\exp P(A_1, A_2)$ with $P(A_1, A_2) \in \mathcal{L}$, then the decomposition $P = \sum_{n \geq 1} P_n$ into homogeneous elements $P_n$ of degree $n$ satisfies $P_1 = 0$, i.e. $P \in \mathcal{L}'$, $P_2 = (1/24)[A_1, A_2]$, $P_3 = a_3([A_1, [A_1, A_2]] - [A_2, [A_2, A_1]])$, where $a_3 \in K$ is arbitrary.

Given any tensor category $\mathcal{T}$ with an infinitesimal braiding $t$, we can construct a ribbon category $\mathcal{T}[[h]]$, as follows. It has the same objects as $\mathcal{T}$, but a morphism $X \to Y$ in $\mathcal{T}[[h]]$ is a formal series $\sum_{n \geq 0} f_n h^n$, where $f_n : X \to Y$ is a morphism of $\mathcal{T}$. The tensor product of objects is the same as in $\mathcal{T}$, and the tensor product of morphisms is the extension to $K[[h]]$ of the one in $\mathcal{T}$, with the non-trivial associator

$$\phi_{X,Y,Z} = \Phi(h t_{X,Y}, h t_{Y,Z}). \tag{3.22}$$

To deform the duality, one needs the remark after (3.8), (3.9) with $a_0$, $b_0$ the duality morphisms of the strict category $\mathcal{T}$. The morphisms $\lambda_X$ and $\lambda_Y$ are invertible, for any formal power series in $h$ beginning with 1 is invertible. The braiding is given by

$$R_{X,Y} = \sigma_{X,Y} \exp\left(\frac{h}{2} t_{X,Y}\right), \tag{3.23}$$

and the ribbon structure by

$$v_X = \exp\left(-\frac{h}{2} \gamma_X\right), \tag{3.24}$$

where $\gamma_X = -(\mathrm{id}_X \otimes a\, t_{A,X})(b \otimes \mathrm{id}_X)$ plays the role of the quadratic Casimir operator and $A$ is a left dual of $X$. By the additivity property (3.14) of $t$, we have

$$2\, t_{X,Y} = \gamma_{X \otimes Y} - \gamma_X \otimes \mathrm{id}_Y - \mathrm{id}_X \otimes \gamma_Y, \tag{3.25}$$

and the relation (3.10) follows.

Now we can define a generalized Reshetikhin-Turaev functor $F : \mathcal{R} \to \mathcal{T}[[h]]$, which restricted to closed ribbon graphs gives an invariant of oriented framed links. We choose first a fixed object $X_\downarrow$ in $\mathcal{T}[[h]]$, and a left dual $X_\uparrow$ of $X_\downarrow$. We put $F(\langle x, \downarrow \rangle) = X_\downarrow$, $F(\langle x, \uparrow \rangle) = X_\uparrow$ and extend the definition of $F$ to all objects of $\mathcal{R}$ by requiring $F(w \square w') = F(w) \otimes F(w')$. To define $F$ on morphisms, we require that $F$ preserves $\phi$, the braiding $R$, the family of morphisms $a$ and $b$ which define duality, and the ribbon structure $v$.

# 4  The category of diagrams

A chord diagram $D$ is a pair $(X, C)$, where $X$ is a set of lines and circles and $C$ is a set of chords connecting pairs of points in $X$. The precise definition is as follows: $X$ is a compact oriented, piecewise smooth one-dimensional submanifold of $\mathbb{R}^2 \times [0,1] = \{(x, y, t) | 0 \le t \le 1\}$ such that:

(i) its connected components $X_j$, $j = 1, \ldots, m$, are homeomorphic to $[0, 1]$ or $S^1$,

(ii) $\partial X = N_0 \cup N_1$, where $N_i = \partial X \cap E_i$, $E_i = \{(x, 0, i) \,|\, x \in \mathbb{R}\}$,

(iii) for each $x \in N_i$, $i = 0, 1$, the function $t$ restricted to the tangent line to $X$ at $x$ is not constant, i.e. the tangent line is not horizontal.

A chord on $X$ is a subset $\{x, y\} \subset X \backslash \partial X$ of two elements $x \ne y$, and $C$ is a finite set of disjoint chords on $X$, possibly empty. $X$ is called the support of the diagram. If $\text{Card}(C) = n$ we shall say that the diagram $D = (X, C)$ is of order $n$. A point $x \in X$ such that $\{x, y\} \in C$ for some $y \in X$ is called a vertex. Chord diagrams are represented as in fig. 13.

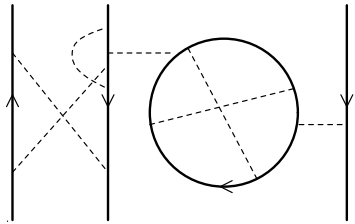

Figure 13:

We identify two diagrams if they are related by a diffeomorphism which sends the lines $E_0$ and $E_1$ into themselves and conserves the orientation of $X$, $E_0$ and $E_1$. Note that in particular this implies that over and under-crossings are identified in chord diagrams, see fig. 14.

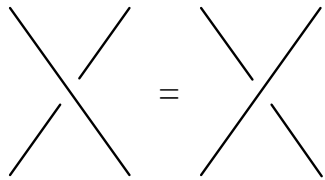

Figure 14:

Let $K$ be a field of characteristic 0, and $D^{(n)}$ be the $K$-vector space spanned by the diagrams of order $n$. The space of Bar-Natan diagrams of order $n$ is $B^{(n)} = D^{(n)}/R^{(n)}$, where $R^{(n)}$ is the subspace of $D^{(n)}$ spanned by the linear combinations of 4 diagrams defined in fig. 15. Note that the 4 terms of fig. 15 stand for arbitrary diagrams $D_1, \ldots, D_4$ which are equal except for the parts shown there.

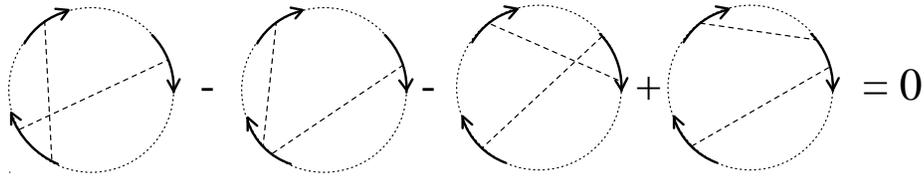

Figure 15:

Now we are ready to define the category $\mathcal{D}_K$ of diagrams. Its objects are finite sequences $S = (\alpha_1, \alpha_2, \ldots, \alpha_k)$, where $\alpha_j \in \{\uparrow, \downarrow\}$, $j = 1, \ldots, k$, including the empty sequence $\emptyset$. A Bar-Natan diagram with support $X$ is a morphism from $S_0$ to $S_1$, the $S_i$ being defined as follows: use the fact that $E_i$ is an ordered set, to build a sequence of unit tangent vectors $\tilde{S}_i = (u_1, u_2, \ldots, u_k)$, where $u_1$ is the tangent vector to $X$ at the smallest $x \in N_i$, $u_2$ is the next tangent vector to $X$ along $E_i$, and so on. Then $S_i$ is given by the projection of $\tilde{S}_i$ on the vertical axis $t$.

The composition of morphisms is defined as follows: if $b : S_0 \to S_1$ and $b' : S'_0 \to S'_1$ are morphisms with $S_1 = S'_0$ defined by the diagrams $(X, C)$ and $(X', C')$, we first translate $(X', C')$ in $\mathbb{R}^2 \times [1, 2]$, then glue $N_1$ to $N'_0$ and finally scale $t$ for $0 \leq t \leq 2$ by a factor $1/2$. This defines a chord diagram $(X'', C'')$ and a morphism $b'' = b'b : S_0 \to S'_1$. This is illustrated on fig. 16.

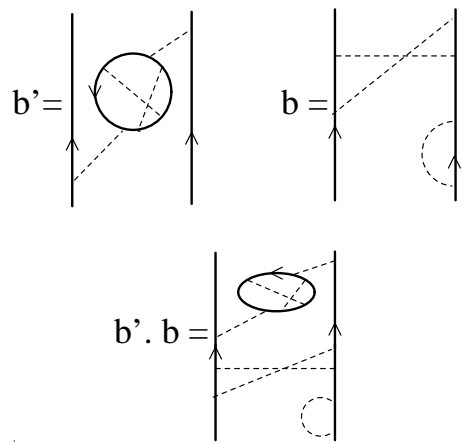

Figure 16:

The identity morphism is represented by the diagram of order 0 on fig. 17.

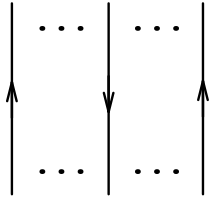

Figure 17:

This category of diagrams $\mathcal{D}_K$ is a tensor category. The monoidal structure $\otimes$ is defined on objects as concatenation of sequences, the unit object being $\emptyset$. For morphisms $b \otimes b'$ is juxtaposition, see fig. 18. The symmetric braiding $\sigma_{S,S'}$ is given by either one of the diagrams of fig. 14.

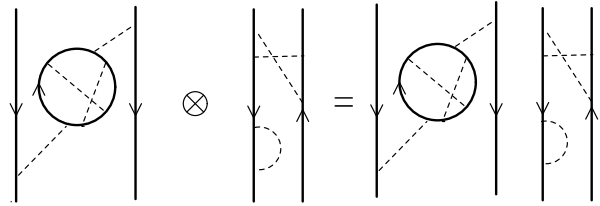

Figure 18:

For $\alpha, \alpha' \in \{\uparrow, \downarrow\}$, let $\Omega_{\alpha,\alpha'}$ be the diagram of fig. 19, with $\alpha$ and $\alpha'$ given by the orientations (not shown on the figure) of the two lines. The category $\mathcal{D}_K$ has an infinitesimal braiding $t$ defined by $t_{\alpha,\alpha'} = \pm \Omega_{\alpha,\alpha'}$, where the sign is + or - according to whether $\alpha = \alpha'$ or not. We denote by $t_{ij}$, with $i, j \in \{1, \ldots, n\}$ any diagram of degree one, whose support has $n$ connected components which are parallel vertical lines, such that it reduces to a diagram $t_{\alpha,\alpha'}$ if all lines but those labeled $i$ and $j$ are deleted.

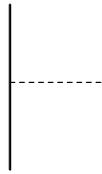

Figure 19:

For any tensor category $\mathcal{T}$ over $K$ with an infinitesimal braiding $t$, and any object $X_\downarrow$ in $\mathcal{T}$ with a left dual $X_\uparrow$, there is a functor of tensor categories $W_\mathcal{T} : \mathcal{D}_K \to \mathcal{T}$ defined by $W_\mathcal{T}(\downarrow) = X_\downarrow$, $W_\mathcal{T}(\uparrow) = X_\uparrow$ and $W_\mathcal{T}(t_{\alpha,\alpha'}) = t_{X_\alpha, X_{\alpha'}}$, for $\alpha, \alpha' \in \{\uparrow, \downarrow\}$. This functor defines a weight system (Feynman rules) on $\mathcal{D}_K$. The Reshetikhin-Turaev functor $F_U : \mathcal{R} \to \mathcal{D}_K[[h]]$ defined in the previous section is called the universal Vassiliev invariant. For any Reshetikhin-Turaev functor $F : \mathcal{R} \to \mathcal{T}[[h]]$, $F = W_\mathcal{T} \circ F_U$, because $\mathcal{D}_K$ is a free [19] tensor category with infinitesimal braiding.

From several sources [18, 20, 13] we can extract the next theorem, which enables the computation of $F_U(L)$ for any framed link (closed ribbon graph) $L$. Before stating it, we need a definition: let $\mathrm{id}_k$ be the identity morphism of an object $S$ in $\mathcal{D}_K$ which is a sequence of $k$ arrows. We consider $\mathrm{id}_k$ as a standard morphism in $\mathcal{D}_K[[h]]$. For any morphism $f$ in $\mathcal{D}_K[[h]]$, define $f_{(k)}$ recursively by $f_{(0)} = f$, $f_{(k)} = f_{(k-1)} \otimes \mathrm{id}_1$, and put $f_{j,k} = (\mathrm{id}_j \otimes f)_{(k)}$.

**Theorem 1** *The functor $F_U$ takes on the elementary standard ribbon graphs the values*

$$F_U(X_{i,n}^\pm) = (\Phi^{(i)})^{-1} \sigma_{i,i+1} \exp(\pm \frac{h}{2} t_{i,i+1}) \Phi^{(i)} \tag{4.1}$$

$$F_U(\bigcap_{i,n}) = a_{i-1,n-i-1} \Phi^{(i)} \tag{4.2}$$

$$F_U(\bigcap_{i,n}) = (\Phi^{(i)})^{-1} b_{i-1,n-i-1} \tag{4.3}$$

*where*

$$\Phi^{(i)} = \Phi(h \sum_{j=1}^{i-1} t_{ji}, h t_{i,i+1}). \tag{4.4}$$

In fact $F_U$ depends on the choice of a solution $\Phi$ to (3.18-3.20). What is remarkable is that for closed graphs $L$ (framed links), $F_U(L)$ is independent of the choice of $\Phi$ [21]. This implies that the coefficients of the universal Vassiliev invariant of links are rational.

We denote by $\mathcal{A}(X)$ the space of all chord diagrams with support $X$. The space $\mathcal{A} = \mathcal{A}(S^1)$ has the structure of a commutative algebra, with the multiplication defined by means of the connected sum of the two circles [9, 10]. If $X$ has $m$ connected components $X_j$, $1 \leq j \leq m$, then for each value of $j$ we can define an $\mathcal{A}$-module $\mathcal{A}_j(X)$, which is isomorphic to $\mathcal{A}(X)$ as a vector space, and for which the action of $\mathcal{A}$ is given by the connected sum $S^1 \# X_j$, where $S^1$ is the circle of $\mathcal{A} = \mathcal{A}(S^1)$. This is illustrated in fig. 20. The 4-term relation of fig. 15 implies that the result is independent of the point of insertion of the circle.

For further convenience we define $\Theta \in \mathcal{A}$ as the unique chord diagram on $S^1$ of degree 1, shown on fig. 21.

## 5 The Kontsevich integral

Let $G$ be a standard ribbon graph, i.e. a morphism of $\mathcal{R}$ from $w$ to $w'$, where $w$ and $w'$ are two standard words. Remember that $G$ is an equivalence class of ribbons which are related by an isotopy of $\mathbb{C} \times [0,1] = \{(z,t) \mid 0 \leq t \leq 1\}$. We choose a particular element $\hat{G}$ in this class, such that $t$ is a Morse function on $\hat{G}$, i.e. for a given value $t = t_0$ there is at most one extremum of $\hat{G}$. Each connected component $\hat{G}_j$ is the image of an embedding $\iota : G_j^0 \times [0,1] \to \mathbb{C} \times [0,1]$, where $G_j^0 = [0,1]$ or $S^1$. We assume that $\hat{G}$ is contained in the plane $\text{Im}(z) = 0$ except for small neighbourhoods around the crossings of figs. 4 and 5, where only the framing vector $\iota(x,1) - \iota(x,0)$ is contained in the plane $\text{Im}(z) = 0$. In other words, we use the blackboard framing. Let $X_j(\hat{G})$ be the curve $\iota(G_j^0 \times \{0\})$, $X(\hat{G}) = \bigcup_j X_j(\hat{G})$. Let $h$ be a formal variable, $\hbar = h(2\pi i)^{-1}$, $\epsilon > 0$ and

$$Z_\epsilon(\hat{G}) = \sum_{n=0}^{\infty} \hbar^n \int_{\substack{t_{\min} < t_1 < \cdots < t_n < t_{\max} \\ |z_i - z_i'| > \epsilon}} \sum_{\text{pairings } P = \{(z_i, z_i')\}} (-1)^{\#P_\uparrow} D(\hat{G}, P) \prod_{i=1}^{n} \frac{dz_i - dz_i'}{z_i - z_i'} \quad (5.1)$$

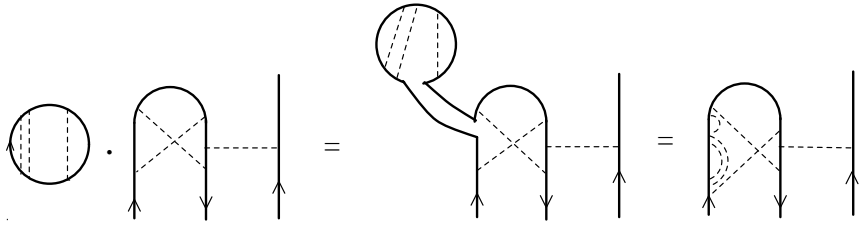

Figure 20:

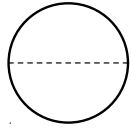

Figure 21:

Here $t_{\min}$ and $t_{\max}$ are the minimal and maximal value of $t$ on $X(\hat{G})$, a pairing $P$ is a choice of $n$ unordered pairs $(z_i, z_i')$, such that for $1 \leq i \leq n$, $(z_i, t_i)$ and $(z_i', t_i)$ are distinct points on $X(\hat{G})$, $\#P_\uparrow$ is the number of vertices $(z_i, z_i')$ of $P$ at which $X(\hat{G})$ is oriented upwards. The coordinates $z_i$ and $z_i'$ are considered as functions of $t_i$, and integration is over the subset of the $n$-simplex $t_{\min} < t_1 < \cdots < t_n < t_{\max}$ defined by the conditions $|z_i(t_i) - z_i'(t_i)| > \epsilon$. $D(\hat{G}, P)$ is the chord diagram of degree $n$ with the support $X(\hat{G})$ and the set of chords $C$ defined by the pairing $P$. (fig.22).

Now we define the regularized Kontsevich integral as:

$$Z(\hat{G}) = \lim_{\epsilon \to 0} \prod_{j=1}^{m} \epsilon^{\hbar(n_j^+ - n_j^-)\Theta_j} \cdot Z_\epsilon(\hat{G}) \quad (5.2)$$

where $n_j^\pm$ are the number of critical points (maxima and minima) of the Morse function $t$ on the component $X_j(\hat{G})$, and $\Theta_j$ denotes the diagram $\Theta$ acting on the latter. (Notice that $n_j^+ - n_j^- = 0$ if $j$ corresponds to a circle.) Later we will show that a slightly modified version of $Z(\hat{G})$ is invariant under isotopies of ribbons.

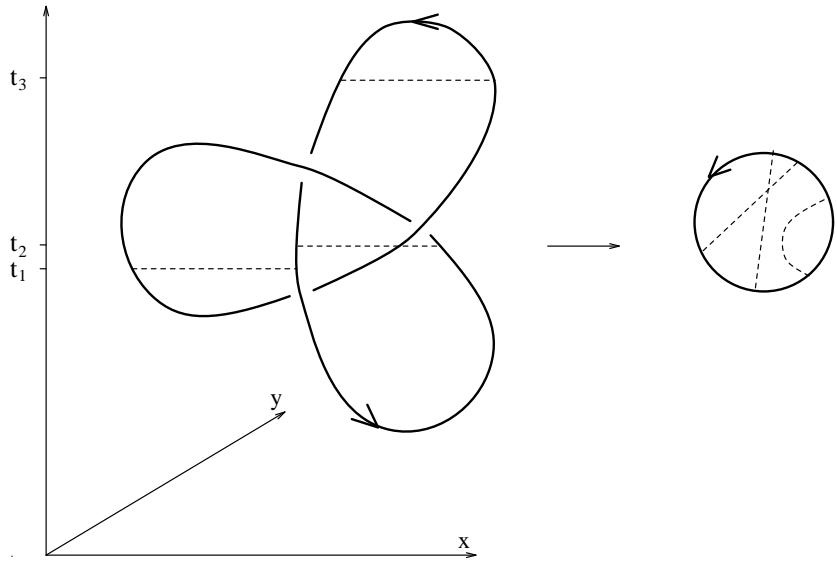

Figure 22:

We recover the usual definition of the Kontsevich invariant if we impose the additional relation $\Theta = 0$, i.e if every diagram having an isolated cord is set equal to zero. This condition implements the framing independence in the original Kontsevich integral [10, 9].

**Theorem 2** *The regularized Konsevitch integral is well-defined and multiplicative:*

$$Z(\hat{G}\,\hat{G}') = Z(\hat{G})\,Z(\hat{G}'). \tag{5.3}$$

*Proof.* A pairing $P''$ of $\hat{G}\,\hat{G}'$ can be partitioned into a disjoint union of pairings $P, P'$ of $\hat{G}$ and $\hat{G}'$, such that $D(\hat{G}\,\hat{G}', P'') = D(\hat{G}, P)\,D(\hat{G}', P')$. Let

$$f = (-1)^{\#P_\uparrow} D(\hat{G}\,\hat{G}', P'') \prod_{i=1}^{n} f_i(t_i), \tag{5.4}$$

where

$$f_i(t_i) = \begin{cases} \frac{\partial}{\partial t_i} \log(z_i - z_i'), & \text{if } |z_i - z_i'| > \epsilon \\ 0 & \text{otherwise.} \end{cases} \tag{5.5}$$

The formula

$$\int_{a<t_1<\ldots<t_n<b} dt_1 \ldots dt_n\, f = \sum_{p=0}^{n} \int_{a<t_1<\ldots<t_p<c} dt_1 \ldots dt_p \int_{c<t_{p+1}<\ldots<t_n<b} dt_{p+1} \ldots dt_n\, f \tag{5.6}$$

implies that $Z_\epsilon(\hat{G}\,\hat{G}') = Z_\epsilon(\hat{G})\,Z_\epsilon(\hat{G}')$. The action of each factor $\Theta_j$ in (5.2) through the connected sum is independent of its point of insertion on the diagram $D(\hat{G}, P)$, thus we can "move" it along the support until it reaches an extremum. By the definition of the Morse function $t$ we can decompose $\hat{G}$ into a product of standard graphs each of which contains only one extremum. Therefore the theorem is proved if we can show that $Z$ converges for such a graph. This is done in lemma 1 below. □

The main ingredient in the construction of the Kontsevich invariant $Z(\hat{G})$, as we now explain, appears when $\hat{G}$ contains no extremum, so that it becomes a braid. Consider a path $\gamma : [0,1] \to V_m$, where

$$V_m = \{(z_1, \ldots, z_m) \in \mathbb{C}^m \mid z_i \neq z_j, i \neq j\}. \tag{5.7}$$

Let $\gamma_i(t) = z_i(t)$, $t \in [0,1]$ be the $i$-th component of this path. Construct a ribbon graph $\hat{G}^\gamma$ out of $\gamma$, such that its $i$-th connected component $\hat{G}_i^\gamma : [0,1] \times [0,1] \to \mathbb{C} \times [0,1]$ is given by

$$\hat{G}_i^\gamma(t, u) = (\gamma_i(t) + u\delta, t), \tag{5.8}$$

where $u \in [0, 1]$, and $\delta \in \mathbb{R}$ is a constant.

Suppose first that $\gamma$ is a closed path. Let $\Omega_{ij}$ be the chord diagram of degree one, with support $\bigcup_j \{(\gamma_j(t), t) \mid t \in [0,1]\}$, and its chord connecting the components labeled by $i$ and $j$. Define the abstract Knizhnik-Zamolodchikov connection with values in the algebra $\mathcal{B}_m$ generated by the $\Omega_{ij}$, $1 \leq i, j \leq m$, as the 1-form on $V_m$:

$$\omega = \hbar \sum_{i<j} s_i s_j \Omega_{ij} \frac{dz_i - dz_j}{z_i - z_j} \tag{5.9}$$

where $s_i = \pm 1$ according to whether the orientation of the $i$-th component is downwards or upwards. It is easy to see that

$$Z(\hat{G}^\gamma) = \overleftarrow{\exp} \int_0^1 \gamma^* \omega. \tag{5.10}$$

Let us quickly recall some well-known properties [22] of $Z(\hat{G}^\gamma)$. Let $\gamma'$ be another closed path in $V_m$, with $\gamma'(1) = \gamma(0)$, and let $\gamma' \cdot \gamma$ denote their product. Then we have $Z(\hat{G}^{\gamma' \cdot \gamma}) = Z(\hat{G}^{\gamma'})\, Z(\hat{G}^\gamma)$. Set $t_{ij} = s_i s_j \Omega_{ij}$. The four-term relation satisfied by diagrams (fig. 15) means that the generators $t_{ij}$ of $\mathcal{B}_m$ satisfy (3.16). The KZ connection $\omega$ is closed, $d\omega = 0$, and the relation (3.16) is equivalent to $\omega \wedge \omega = 0$, so that $\omega$ is flat: $F_\omega = d\omega + \omega \wedge \omega = 0$. This implies that $Z(\hat{G}^\gamma)$ only depends on the homotopy class of $\gamma$, so that the map $\gamma \mapsto Z(\hat{G}^\gamma)$ defines a representation $\pi_1(V_m, *) \to \mathcal{B}_m$, where $* = \gamma(0) = \gamma(1)$ is a fixed basepoint. Since $\pi_1(V_m, *)$ is the pure braid group on $m$ strands $P_m$ [23], the Kontsevich integral $Z(\hat{G}^\gamma)$ gives a representation of $P_m$ in the algebra $\mathcal{B}_m$.

Let now $\gamma$ be an open path in $V_m$, with $\gamma(1) = \sigma \cdot \gamma(0)$, with $\sigma \in S_m$ a permutation of the $m$ coordinates. This path $\gamma$ defines an element of the full braid group $B_m = \pi_1(V_m/S_m, *)$, and

$$Z(\hat{G}^\gamma) = D_\sigma \overleftarrow{\exp} \int_0^1 \gamma^* \omega, \tag{5.11}$$

where $D_\sigma$ is a diagram of order zero in which the components get permuted according to $\sigma$. See fig. 14 for the case of the non-trivial element of $S_2$. Thus the formula (5.11) gives a representation of the full braid group $B_m$ in the semidirect product of $\mathcal{B}_m$ with $S_m$, where the symmetric group $S_m$ is identified with the group of order zero diagrams $D_\sigma$ with $m$ components.

We can also write $Z(\hat{G}^\gamma)$ as an ordered exponential in the case where $\gamma$ is an open path contained in a simply-connected subset of $V_m$, such as the $m$-simplex $\Delta_m = \{(z_1, \ldots, z_m) \mid z_i \in \mathbb{R},\ 0 < z_1 < z_2 < \cdots < z_m < 1\}$. In this case the formula (5.10) also holds. We warn the readers, that although in this case $\gamma$ is a path in the space $\Delta_m$ which is homotopically trivial, in general $Z(\hat{G}^\gamma) \neq D_1$ (here 1 is the identity permutation). Examples will follow shortly.

Returning to the general case, to conclude this brief reminder about the consequences of the flatness of $\omega$, one can show that for any open path $\gamma$ of $V_m$, the integral $Z(\hat{G}^\gamma)$ depends only on the homotopy class of $\gamma$ relative to the endpoints.

The morphisms of $\mathcal{R}$ are generated by the crossings $X_{m,n}^\pm$ and the graphs $\bigcap_{m,n}$ and $\bigcup_{m,n}$ of figs. 4-7 containing a local minimum or maximum of the function $t$, and $m - 1$ (resp. $n - m - 1$) vertical lines on the left (right).

**Lemma 1** $Z(\bigcap_{m,n})$ and $Z(\bigcup_{m,n})$ converge.

*Proof.* In the following, we use the notation $Z(\gamma)$ instead of $Z(\hat{G}^\gamma)$, for $\gamma$ a path in $V_m$, which is not supposed to be closed. We start by computing $Z(\gamma)$ when $\gamma$ is a path with 2 components of opposite orientations. We parametrize the components as follows:

$$z_i(t) = x_i + (y_i - x_i)t,\ \ 0 \leq t \leq 1, \tag{5.12}$$

for $i = 1, 2$, with $x_i, y_i \in \mathbb{R}$, $x_1 < x_2$ and $y_1 < y_2$. Choose an $\epsilon > 0$ such that $\epsilon \leq \min\{x_2 - x_1, y_2 - y_1\}$. Then eq. (5.1) or (5.10) leads to

$$Z_\epsilon(\gamma) = Z(\gamma) = \exp\left(-\overline{\Omega}_{12} \log \frac{y_1 - y_2}{x_1 - x_2}\right) = \left(\frac{y_1 - y_2}{x_1 - x_2}\right)^{-\overline{\Omega}_{12}}. \tag{5.13}$$

Here we have used the notation $\overline{\Omega}_{ij} = \hbar \Omega_{ij}$. Suppose $x_2 - x_1 > y_2 - y_1$, take now $\epsilon = y_2 - y_1$. Then

$$\lim_{\epsilon \to 0} \epsilon^{\overline{\Omega}_{12}} Z_\epsilon(\gamma) \tag{5.14}$$

exists, and the same is true in the other case $\epsilon = x_2 - x_1 < y_2 - y_1$, namely

$$\lim_{\epsilon \to 0} Z_\epsilon(\gamma) \, \epsilon^{-\overline{\Omega}_{12}}. \tag{5.15}$$

Consider now $\gamma_{m,n}$, the same path with $m - 1$ (resp. $n - m - 1$) vertical lines on the left (right) and the two center lines as in $\gamma$. The parametrization we choose is $z_i = \text{const} \in \mathbb{R}$ for $i \neq m, m + 1$ and $z_i$ as in (5.12) for $i = m, m + 1$. We will now prove that the limits

$$\lim_{\epsilon \to 0} \epsilon^{\overline{\Omega}_{m,m+1}} \, Z_\epsilon(\gamma_{m,n}) \tag{5.16}$$

and

$$\lim_{\epsilon \to 0} Z_\epsilon(\gamma_{m,n}) \, \epsilon^{-\overline{\Omega}_{m,m+1}} \tag{5.17}$$

exist, where $\epsilon = \min\{x_{m+1} - x_m, y_{m+1} - y_m\}$, and $\max\{x_{m+1} - x_m, y_{m+1} - y_m\}$ is fixed. The KZ connection can be written as

$$\omega = -\overline{\Omega}_{m,m+1} \frac{dz_m - dz_{m+1}}{z_m - z_{m+1}} + \omega', \tag{5.18}$$

where

$$\omega' = \sum_{k \neq m, m+1} s_m s_k \left( \overline{\Omega}_{m,k} \frac{dz_m}{z_m - z_k} - \overline{\Omega}_{m+1,k} \frac{dz_{m+1}}{z_{m+1} - z_k} \right). \tag{5.19}$$

Using the factorization formula (2.10), we find

$$Z(\gamma_{m,n}) = g(y_m - y_{m+1}) \, \overleftarrow{\exp} \int_0^1 \left( g(z_m - z_{m+1})^{-1} \omega' \, g(z_m - z_{m+1}) \right), \tag{5.20}$$

where

$$g(z) = \left( \frac{z}{x_m - x_{m+1}} \right)^{-\overline{\Omega}_{m,m+1}}. \tag{5.21}$$

Therefore, the relation (3.16) implies that

$$g(y_m - y_{m+1})^{-1} Z(\gamma_{m,n}) = \overleftarrow{\exp} \int_0^1 (\omega' + \omega'') \tag{5.22}$$

where

$$\omega'' = \sum_{p=1}^\infty \frac{1}{p!} \log^p \left( \frac{z_m - z_{m+1}}{x_m - x_{m+1}} \right) \text{ad}^p(\overline{\Omega}_{m,m+1}) \sum_{k \neq m, m+1} s_m s_k \overline{\Omega}_{mk} \left( \frac{dz_m}{z_m - z_k} - \frac{dz_{m+1}}{z_{m+1} - z_k} \right), \tag{5.23}$$

and to prove that (5.16) exists, it is enough to show that

$$I_p = \int_0^1 dt \, \log^p(z_{m+1} - z_m) \left( \frac{\dot{z}_m}{z_m - z_k} - \frac{\dot{z}_{m+1}}{z_{m+1} - z_k} \right) \tag{5.24}$$

converges when $y_m \to y_{m+1}$, for any positive integer $p$. Here we have adopted the notation $dz/dt = \dot{z}$. Writing

$$I_p = \int_0^1 dt \, \log^p(z_{m+1} - z_m) \left( \frac{\dot{z}_m(z_{m+1} - z_m)}{(z_m - z_k)(z_{m+1} - z_k)} - \frac{\dot{z}_{m+1} - \dot{z}_m}{z_{m+1} - z_k} \right), \tag{5.25}$$

and noting that $(z_{m+1} - z_m) \log^p(z_{m+1} - z_m) \to 0$ as $z_{m+1} - z_m \to 0$, we see that the first term converges. As for the second term, we put

$$(\dot{z}_{m+1} - \dot{z}_m) \log^p(z_{m+1} - z_m) = \frac{d}{dt} F_p(z_{m+1} - z_m), \tag{5.26}$$

where

$$F_p(z) = \int_{z_0}^z dx \, \log^p x, \tag{5.27}$$

and observe that the function $F_p(z)$ is defined at $z = 0$. Thus, integrating by parts proves that

$$\int_0^1 \frac{dt \, dF_p(z_{m+1} - z_m)/dt}{z_{m+1} - z_k} \tag{5.28}$$

converges. This concludes the proof that the limit (5.16) exists. Similarly, one can show the existence of (5.17).

Notice that since the KZ connection is flat, (5.16) and (5.17) do not depend on the details of the path $\gamma_{m,n}$, but only on its endpoints. From the definition of the regularized integral (5.2), we see that for $\epsilon = \min\{x_{m+1} - x_m, y_{m+1} - y_m\}$, keeping $\max\{x_{m+1} - x_m, y_{m+1} - y_m\}$ fixed,

$$Z(\cap_{m,n}) = \lim_{\epsilon \to 0} \cap'_{m,n} \epsilon^{\overline{\Omega}_{m,m+1}} Z_\epsilon(\gamma_{m,n}) = \lim_{\epsilon \to 0} (\epsilon^{\hbar \Theta_m} \cdot \cap'_{m,n}) Z_\epsilon(\gamma_{m,n}), \tag{5.29}$$

$$Z(\cup_{m,n}) = \lim_{\epsilon \to 0} Z_\epsilon(\gamma_{m,n}) \, \epsilon^{-\overline{\Omega}_{m,m+1}} \cup'_{m,n} = \lim_{\epsilon \to 0} Z_\epsilon(\gamma_{m,n}) \, (\epsilon^{-\hbar \Theta_m} \cdot \cup'_{m,n}), \tag{5.30}$$

where $\cap'_{m,n}$ and $\cup'_{m,n}$ are diagrams in $\mathcal{D}_K$ of order zero with supports given by $\cap_{m,n}$, $\cup_{m,n}$, such that the extremities of the line containing the maximum (resp. minimum) are $y_m$ and $y_{m+1}$ ($x_m$ and $x_{m+1}$). □

Now we want to show that (5.29) and (5.30) are independent of the location of the extremum. We give the proof for a maximum. For this we proceed in two steps. The first step is illustrated in fig. 23. The figure represents the expression $\cap'_{m,n} \epsilon^{\overline{\Omega}_{m,m+1}} Z_\epsilon(\gamma_{m,n})$. Using flatness one replaces $\gamma_{m,n}$ by a path whose top part $T$ has the lines $z_m$ and $z_{m+1}$ satisfying $|z_m - z_{m+1}| = \epsilon$. In the second step, we show that the contribution of this top part is trivial: $Z(T) = T$, the diagram with no chords, therefore $T$ can be erased. To prove that the second step is justified, we observe that if there exists a constant $C > 0$ independent of $\epsilon$ such that $|z_m - z_k| \geq C$, for all $k \neq m, m+1$, then the integral $I_p$ corresponding to $T$ – defined in (5.24) and written conveniently in (5.25) – tends to zero when $\epsilon$ goes to zero.

Note that these arguments apply not only to $\bigcap_{m,n}$, but to all the graphs $A\,\square_{\mathrm{st}}\bigcap_{1,2}\square_{\mathrm{st}}B$, where $A,B$ are arbitrary standard ribbon graphs. They show that the maxima and minima can be moved freely without changing the value of $Z$, except for the two configurations shown on fig. 24 and 25, where there is no constant $C$ satisfying the conditions. In the case of fig. 24 this is not a failure, but a nice feature of $Z$, since we want to get an invariant of ribbons. One cannot remove the parts of a ribbon graph which look like fig. 25 without changing the value of $Z$, because the diagrams $\bigcap'_{1,2}$ and $\bigcup'_{1,2}$ are not the morphisms $a$ and $b$ of $\mathcal{D}_K[[h]]$, they satisfy only the weak properties (3.8) and (3.9). However, if we define $\hat{Z}(\hat{G})$ by

$$\hat{Z}(\bigcap_{m,n}) = \mu^{-1}\cdot Z(\bigcap_{m,n}), \tag{5.31}$$

where $\mu = Z(U)$ and $U$ is the diagram of fig. 26, acting on the component of $\bigcap_{m,n}$ carrying the maximum, and $\hat{Z}(\hat{G}) = Z(\hat{G})$ for the other elementary graphs $\hat{G}$, then $\hat{Z}$ is invariant under insertion or removal of the subgraph of fig. 25. The flatness of the KZ connection implies that $\hat{Z}$ is invariant under an isotopy which fixes the extrema. Thus we get:

**Theorem 3** $\hat{Z}(\hat{G})$ is an isotopy invariant of ribbon graphs.

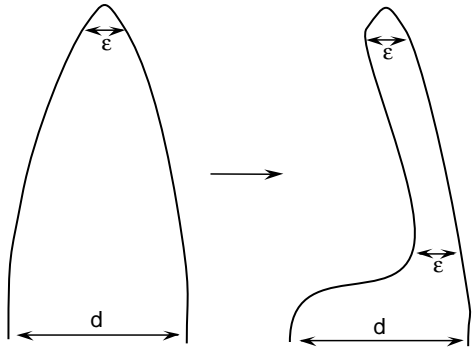

Figure 23:

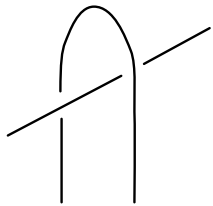

Figure 24:

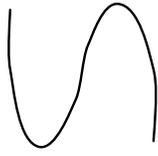

Figure 25:

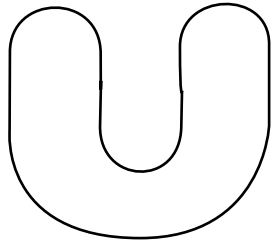

Figure 26:

## 6 The distancing operator

In this section we introduce the distancing operator, which is the renormalized Kontsevich integral of a trivial braid with a pair of consecutive strands moving away from each other to infinity. Let $a \in \mathbb{R}$, $\lambda > 0$. We consider a path $\eta_i^{(n)} : [0,1] \to \Delta_n$, such that

$$
\begin{aligned}
\eta_i^{(n)}(1) &= (z_1, \ldots, z_n), \\
\eta_i^{(n)}(0) &= (z_1 - a, \ldots, z_i - a, z_{i+1} + \lambda - a, \ldots, z_n + \lambda - a).
\end{aligned} \quad (6.1)
$$

(The strands $i$ and $i+1$ are moving away from each other.) Due to the zero curvature condition satisfied by the KZ connection and the fact that $\Delta_n$ is simply-connected we can choose any path with these endpoints to calculate $Z(\eta_i^{(n)})$. Let us take the following one:

$$
\eta_i^{(n)}(t) = \begin{cases} z_j - a(1-t) & 1 \le j \le i \\ z_j + (1-t)(\lambda - a) & i+1 \le j \le n, \, \lambda > 0. \end{cases} \quad (6.2)
$$

It is useful to remark that $Z(\eta_i^{(n)})$ does not depend on the global translation $a$ because the KZ connection depends only on the differences $z_j(t) - z_i(t)$. We put $g_i^{(n)}(\lambda, z_1, \ldots, z_n) = Z(\eta_i^{(n)})$ and in the following we will not mention the arguments $(z_1, \ldots, z_n)$ if there is no ambiguity. Using the definition of the Kontsevich integral we have:

$$
g_i^{(n)}(\lambda) = \overleftarrow{\exp}(\hbar \int_0^1 dt\, \omega(t)), \quad (6.3)
$$

where

$$
\omega(t) = \sum_{j=1}^{i} \sum_{k=i+1}^{n} \frac{\lambda t_{jk}}{(t-1)\lambda + z_j - z_k} \quad (6.4)
$$

The distancing operator is obtained by sending $\lambda$ to infinity, but to do this a regularization is needed.

**Theorem 4** *Set $X_i^{(n)} = \sum_{j=1}^{i} \sum_{k=i+1}^{n} t_{jk}$.*

1. *The distancing operator*

$$D_i^{(n)}(z_1, \ldots, z_n) = \lim_{\lambda \to +\infty} g_i^{(n)}(\lambda) \, \lambda^{\hbar X_i^{(n)}} \tag{6.5}$$

*is a well-defined function on $V_n$ with values in $\mathcal{B}_n[[h]]$, the algebra of formal power series in $h$ with coefficients in $\mathcal{B}_n$.*

2. *It has the asymptotic behaviour $D_i^{(n)} \sim (z_n - z_1)^{-\hbar X_i^{(n)}}$ in the region*

$$\begin{cases} z_n - z_1 \gg z_n - z_k, & k > i \\ z_n - z_1 \gg z_j - z_1, & j \leq i \end{cases}. \tag{6.6}$$

3. *It satisfies the differential equations*

$$\frac{\partial D_i^{(n)}}{\partial z_j} = \hbar \left( \sum_{\substack{k=1 \\ k \neq j}}^{n} \frac{t_{jk}}{z_j - z_k} \right) D_i^{(n)} - \hbar D_i^{(n)} K_{ij}^{(n)}, \tag{6.7}$$

*where*

$$K_{ij}^{(n)} = \begin{cases} \sum_{\substack{k=1 \\ k \neq j}}^{i} \frac{t_{jk}}{z_j - z_k}, & j \leq i \\ \sum_{\substack{k=i+1 \\ k \neq j}}^{n} \frac{t_{jk}}{z_j - z_k}, & j > i. \end{cases} \tag{6.8}$$

*Proof.* 1. We factorize from $g_i^{(n)}(\lambda)$

$$\overleftarrow{\exp}(\hbar \int_0^1 dt \frac{\lambda X_i^{(n)}}{(t-1)\lambda + z_1 - z_n}) = \left( \frac{z_n - z_1}{\lambda + z_n - z_1} \right)^{\hbar X_i^{(n)}}. \tag{6.9}$$

Thus, using (2.11) and performing the change of variables $x = \lambda(1-t)/(z_n - z_1)$ we get:

$$g_i^{(n)}(\lambda) = \overleftarrow{\exp}(\hbar \int_{\frac{\lambda}{z_n - z_1}}^{0} dx \, \tilde{\omega}(x)) \left( \frac{z_n - z_1}{\lambda + z_n - z_1} \right)^{\hbar X_i^{(n)}}, \tag{6.10}$$

with

$$\tilde{\omega}(x) = (1+x)^{-\hbar X_i^{(n)}} \left( \sum_{\substack{1 \leq j \leq i \\ i+1 \leq k \leq n}} \frac{t_{kj}(1 - u_{kj})}{(x + u_{kj})(1+x)} \right) (1+x)^{\hbar X_i^{(n)}} \tag{6.11}$$

and $u_{kj} = (z_k - z_j)/(z_n - z_1)$. Since

$$\tilde{\omega}(x) = \sum_{p=0}^{+\infty} \frac{\hbar^p}{p!} \left( \sum_{\substack{1 \leq j \leq i \\ i+1 \leq k \leq n}} \mathrm{ad}^p(X_i^{(n)}) \cdot t_{kj} \right) \frac{(1 - u_{kj}) \log^p(1+x)}{(x + u_{kj})(x + 1)}, \tag{6.12}$$

and the integral
$$\int_0^{+\infty} dx \frac{\log^p(1+x)}{(x+u_{kj})(1+x)} \tag{6.13}$$
converges, we deduce that
$$\overleftarrow{\exp}(\hbar \int_{+\infty}^0 dx \tilde{\omega}(x)) \tag{6.14}$$
converges as a formal series in $h$, which means that all the integrals appearing in its expansion converge.

2. The asymptotic region (6.6) is defined by $u_{kj} = 1$, $1 \leq j \leq i$, $i+1 \leq k \leq n$. In this region $\tilde{\omega}(x) = 0$, thus $D_i^{(n)}(z_n - z_1)^{-\hbar X_i^{(n)}} = 1$.

3. The formula (2.8) yields:
$$\frac{\partial}{\partial z_j} g_i^{(n)}(\lambda) = \hbar \int_0^1 dt\, h(1,t) \frac{\partial}{\partial z_j} \left( -\sum_{\substack{1 \leq l \leq i \\ i+1 \leq k \leq n}} \frac{\lambda t_{lk}}{\lambda(1-t) + z_k - z_l} \right) h(t,0), \tag{6.15}$$
where
$$h(x,y) = \overleftarrow{\exp}(\hbar \int_y^x du\, \omega(u)). \tag{6.16}$$
Using
$$\frac{\partial}{\partial z_j} \left( \sum_{\substack{1 \leq l \leq i \\ i+1 \leq k \leq n}} \lambda \frac{t_{lk}}{\lambda(1-t) + z_k - z_l} \right) = \frac{\partial}{\partial t} \left( \sum_{i+1 \leq k \leq n} \frac{t_{jk}}{\lambda(1-t) + z_k - z_j} \right) \tag{6.17}$$
if $j \leq i$, and integrating by parts over $t$, the r.h.s. of (6.15) becomes:
$$\hbar \sum_{i+1 \leq k \leq n} \frac{t_{jk}}{z_j - z_k} g_i^{(n)}(\lambda) - \hbar g_i^{(n)}(\lambda) \sum_{i+1 \leq k \leq n} \frac{t_{jk}}{\lambda + z_j - z_k} \tag{6.18}$$
$$-\hbar \int_0^1 dt\, h(1,t) \left[ \sum_{i+1 \leq k \leq n} \frac{t_{jk}}{\lambda(1-t) + z_k - z_j}, \sum_{\substack{1 \leq j' \leq i \\ i+1 \leq k' \leq n}} \frac{\lambda t_{j'k'}}{\lambda(1-t) + z_{k'} - z_{j'}} \right] h(t,0).$$

Using in the second term the classical Yang-Baxter equation:
$$\left[ \frac{t_{jk}}{z_j - z_k}, \frac{t_{j'k}}{z_{j'} - z_k} \right] + \left[ \frac{t_{jj'}}{z_j - z_{j'}}, \frac{t_{jk}}{z_j - z_k} + \frac{t_{j'k}}{z_{j'} - z_k} \right] = 0, \tag{6.19}$$
and again the derivation property (2.8) we get:
$$\frac{\partial}{\partial z_j} g_i^{(n)}(\lambda) = \hbar \left( \sum_{\substack{k=1 \\ k \neq j}}^n \frac{t_{jk}}{z_j - z_k} \right) g_i^{(n)}(\lambda) - \hbar g_i^{(n)}(\lambda) \left( \sum_{\substack{j'=1 \\ j' \neq j}}^i \frac{t_{jj'}}{z_j - z_{j'}} + \sum_{k=i+1}^n \frac{t_{jk}}{z_j - z_k - \lambda} \right). \tag{6.20}$$

Now multiply on the right this equality by $\lambda^{\hbar X_i^{(n)}}$, and notice that $[X_i^{(n)}, t_{jj'}] = 0$, $1 \leq j, j' \leq i$ and

$$\lim_{\lambda \to \infty} \lambda^{-\hbar X_i^{(n)}} \sum_{k=i+1}^{n} \frac{t_{jk}}{z_j - z_k - \lambda} \lambda^{\hbar X_i^{(n)}} = 0. \tag{6.21}$$

Here the limit means that each term in the $h$ expansion tends to zero. This proves part 3 of the theorem for $1 \leq j \leq i$ (the proof for the case $i+1 \leq j \leq n$ is identical). $\square$

**Remark.** The distancing operator is equivalently defined as being the solution of the differential equations (6.7) on $V_n$, satisfying the asymptotic conditions (6.6).

Let $A$ be a standard ribbon graph such that its top (resp. bottom) endpoints are $(z_1, \ldots, z_i) \in \Delta_i$ (resp. $(z'_1, \ldots, z'_p) \in \Delta_p$), and let $B$ be a standard ribbon graph such that its top (resp. bottom) endpoints are $(z_{i+1}, \ldots, z_n) \in \Delta_{n-i}$ (resp. $(z'_{p+1}, \ldots, z'_q) \in \Delta_{q-p}$). (Here "endpoint" really means the middle point of the intersection of an open ribbon with $\mathbb{R} \times \{0\} \times \{0,1\}$.) We put these graphs side by side such that the top (resp. bottom) endpoints $(z_1, \ldots, z_i, z_{i+1}, \ldots, z_n)$ (resp. $(z'_1, \ldots, z'_p, z'_{p+1}, \ldots, z'_q)$) of $A \square B$ lie in $\Delta_n$ (resp. $\Delta_q$). This is always possible up to a global translation of $B$. The product $A \square B$ is not standard. Let $A \square_{\text{st}} B$ be the same graph as $A \square B$ but with the standard arrangement of parentheses.

**Theorem 5**

$$Z(A \square_{\text{st}} B) = D_i^{(n)}(z_1, \ldots, z_n) \left( Z(A) \otimes Z(B) \right) (D_p^{(q)}(z'_1, \ldots, z'_q))^{-1}. \tag{6.22}$$

*Proof.* Let $(A \square_{\text{st}} B)_\lambda$ be the ribbon graph whose top (resp. bottom) endpoints are

$$(z_1 - \lambda/2, \ldots, z_i - \lambda/2, z_{i+1} + \lambda/2, \ldots, z_n + \lambda/2)$$
$$(\text{ resp. } (z'_1 - \lambda/2, \ldots, z'_p - \lambda/2, z'_{p+1} + \lambda/2, \ldots, z'_q + \lambda/2)). \tag{6.23}$$

Then $A \square_{\text{st}} B$ and $\eta_i^{(n)}(z_1, \ldots, z_n) \, (A \square_{\text{st}} B)_\lambda \, (\eta_p^{(q)}(z'_1, \ldots, z'_q))^{-1}$ are isotopic ribbon graphs (see fig. 27), where the path $\eta_i^{(n)}$ is defined in (6.2), and the same symbol is used here for the associated standard ribbon graph constructed as in (5.8). We also set $(\eta_i^{(n)})^{-1}(t) = \eta_i^{(n)}(1-t)$, $0 \leq t \leq 1$.

Thus

$$Z(A \square_{\text{st}} B) = g_i^{(n)}(\lambda) Z((A \square_{\text{st}} B)_\lambda)(g_p^{(q)}(\lambda))^{-1} \tag{6.24}$$

$$= (g_i^{(n)} \lambda^{\hbar X_i^{(n)}}) \left( \lambda^{-\hbar X_i^{(n)}} Z((A \square_{\text{st}} B)_\lambda) \lambda^{\hbar X_p^{(q)}} \right) (g_p^{(q)} \lambda^{\hbar X_p^{(q)}})^{-1}. \tag{6.25}$$

From the definition of the regularized Kontsevich integral one sees that

$$Z((A \square_{\text{st}} B)_\lambda) = Z(A) \otimes Z(B) + \sum_{n \geq 1} \hbar^n f_{(n)}(\lambda), \tag{6.26}$$

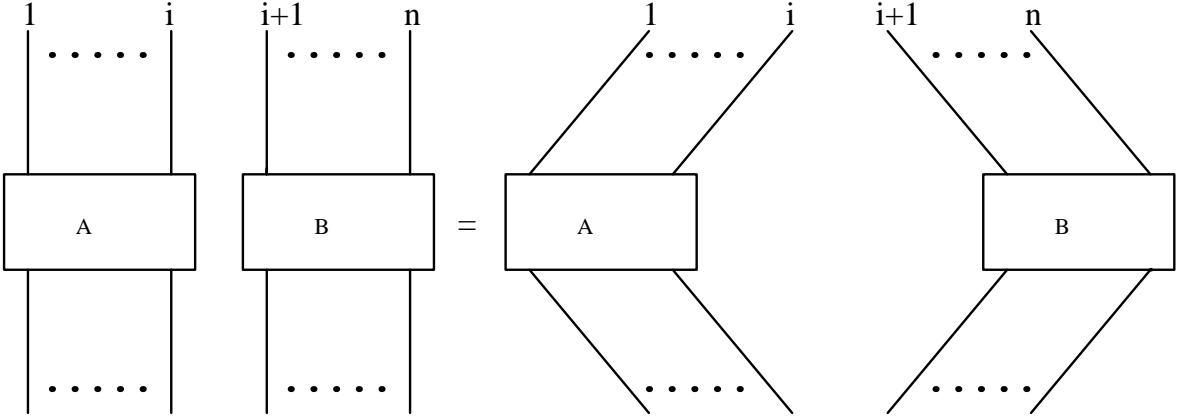

Figure 27:

where $f_{(n)}(\lambda)$ is a finite sum of terms of the type:

$$D(A\Box B, P_{AB}) \int_{t_{\min}}^{t_{\max}} dt \, \frac{\dot{z}_i - \dot{z}_j}{z_i - z_j + \lambda} f(t) \tag{6.27}$$

such that $P_{AB}$ contains at least one pair $(z_i, z_j)$ with $(z_i, t) \in A$, $(z_j, t) \in B$, and $\int_{t_{\min}}^{t_{\max}} f(t)dt$ is convergent. Hence

$$\lim_{\lambda \to \infty} f_{(n)}(\lambda) \, \log^p(\lambda) = 0. \tag{6.28}$$

Moreover the following relation holds : $X_i^{(n)}(Z(A) \otimes Z(B)) = (Z(A) \otimes Z(B)) X_p^{(q)}$. This relation follows from (3.16) and the fact that $D(t_{ik} + t_{i+1,k}) = 0$ (resp. $(t_{ik} + t_{i+1,k})D = 0$) if $D$ is a diagram of degree zero in which the bottom (resp. top) endpoints labeled $i$ and $i+1$ belong to the same connected component (see fig. 28).

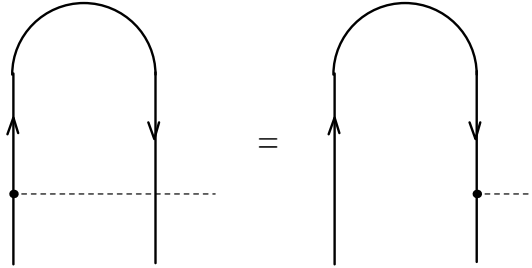

Figure 28:

Therefore

$$\lim_{\lambda \to \infty} \lambda^{-\hbar X_i^{(n)}} Z((A\Box_{\mathrm{st}} B)_\lambda) \lambda^{\hbar X_p^{(q)}} = Z(A) \otimes Z(B), \tag{6.29}$$

and the theorem is proved. □

# 7   The equivalence theorem

The relation between the Drinfeld associator and the distancing operators involves two functions on $V_{i+1}$ with values in $\mathcal{B}_{i+1}[[h]]$:

$$\tilde{g}_{i+1}(z_1,\ldots,z_i,z_{i+1}) = D^{(i+1)}_{i-1} D^{(i-1)}_{i-2} D^{(i-2)}_{i-3} \cdots D^{(2)}_{1} (z_{i+1} - z_i)^{\hbar t_{i,i+1}}, \tag{7.1}$$

$$g_{i+1}(z_1,\ldots,z_i,z_{i+1}) = D^{(i+1)}_{i} D^{(i)}_{i-1} D^{(i-1)}_{i-2} \cdots D^{(2)}_{1}. \tag{7.2}$$

**Theorem 6**

$$(\tilde{g}_{i+1})^{-1} g_{i+1} = \Phi_{KZ}(\hbar X^{(i)}_{i-1}, \hbar t_{i,i+1}) \tag{7.3}$$

*Proof.* The key point is the fact that $\tilde{g}_{i+1}$ and $g_{i+1}$ are solutions of the Knizhnik-Zamolodchikov equations:

$$\frac{\partial g}{\partial z_j} = \hbar \left( \sum_{\substack{k=1 \\ k \neq j}}^{i+1} \frac{t_{jk}}{z_j - z_k} \right) g \tag{7.4}$$

with the following asymptotic behaviour:

$$\tilde{g}_{i+1} \sim (z_2 - z_1)^{\hbar t_{12}} \ldots (z_{i-1} - z_1)^{\hbar X^{(i-1)}_{i-2}} (z_{i+1} - z_1)^{\hbar X^{(i+1)}_{i-1}} (z_{i+1} - z_i)^{\hbar t_{i,i+1}}$$

in the region
$$\begin{cases} z_{i+1} - z_1 \gg z_{i-1} - z_1 \gg \ldots \gg z_2 - z_1 \\ z_{i+1} - z_1 \gg z_{i+1} - z_i; \end{cases} \tag{7.5}$$

$$g_{i+1} \sim (z_2 - z_1)^{\hbar t_{12}} \ldots (z_{i-1} - z_1)^{\hbar X^{(i-1)}_{i-2}} (z_i - z_1)^{\hbar X^{(i)}_{i-1}} (z_{i+1} - z_1)^{\hbar X^{(i+1)}_{i}}$$

in the region   $z_{i+1} - z_1 \gg z_i - z_1 \gg z_{i-1} - z_1 \gg \ldots \gg z_2 - z_1$. \hfill (7.6)

We recall that $X^{(n)}_i = \sum_{j=1}^{i} \sum_{k=i+1}^{n} t_{jk}$. Note that all the factors appearing in the asymptotic behaviour of $g_{i+1}$ and $\tilde{g}_{i+1}$ mutually commute. The relations (7.4), (7.5), and (7.6) are direct consequences of the properties of the distancing operators established in theorem 4.

Any solution $g$ of the KZ equations (7.4) can be expressed in terms of the reduced variables $u_k = (z_k - z_1)/(z_{i+1} - z_1)$, $2 \leq k \leq i$:

$$g(z_1,\ldots,z_i,z_{i+1}) = G(u_2,\ldots,u_i)(z_{i+1} - z_1)^{\hbar X}, \quad X = \sum_{1 \leq j < k \leq i+1} t_{jk}. \tag{7.7}$$

Observe that $X$ is a central element in $\mathcal{B}_{i+1}$. The function $g(z_1,\ldots,z_{i+1})$ satisfies the KZ equations in the variables $z_j$, $j = 1,\ldots,i+1$ if and only if $G(u_2,\ldots,u_i)$ satisfies the equations

$$\frac{\partial G}{\partial u_j} = \hbar \left( \frac{t_{j1}}{u_j} + \frac{t_{j,i+1}}{u_j - 1} + \sum_{2 \leq k \leq i,\, k \neq j} \frac{t_{jk}}{u_j - u_k} \right) G \tag{7.8}$$

for $j = 2, \ldots, i$. Now if we set

$$U_{i+1} = u_2^{-\hbar t_{12}} \ldots u_{i-1}^{-\hbar X_{i-2}^{(i-1)}} G_{i+1}, \tag{7.9}$$

and the same for $\tilde{U}_{i+1}$, $\tilde{G}_{i+1}$, we obtain that $U_{i+1}$ and $\tilde{U}_{i+1}$ are analytic functions in the domain $1 > u_i > u_{i-1} \geq u_{i-2} \geq \ldots \geq u_2 \geq 0$, and they obey the same linear differential equations. Moreover, at the point $u_{i-1} = u_{i-2} = \ldots = u_2 = 0$ it follows from (7.8) that they satisfy the equation

$$\frac{\partial U}{\partial u_i} = \hbar \left( \frac{X_{i-1}^{(i)}}{u_i} + \frac{t_{i,i+1}}{u_i - 1} \right) U \tag{7.10}$$

with the asymptotic behaviour:

$$U_{i+1} \sim (u_i)^{\hbar X_{i-1}^{(i)}} \quad u_i \to 0 \tag{7.11}$$

$$\tilde{U}_{i+1} \sim (1 - u_i)^{\hbar t_{i,i+1}} \quad u_i \to 1. \tag{7.12}$$

Thus, at the point $u_{i-1} = u_{i-2} = \ldots = u_2 = 0$,

$$(\tilde{U}_{i+1})^{-1} U_{i+1} = \Phi_{KZ}(hX_{i-1}^{(i)}, ht_{i,i+1}), \tag{7.13}$$

and by the uniqueness of the solution of differential equations this equality holds in the whole domain $1 > u_i > u_{i-1} \geq u_{i-2} \geq \ldots \geq u_2 \geq 0$. $\square$

We have seen already in section 3 that any closed ribbon graph is a composition of elementary standard graphs $X_{m,n}^{\pm}$, $\cap_{m,n}$ and $\cup_{m,n}$. Recall that $\cap'_{m,n}$ and $\cup'_{m,n}$ are diagrams in $\mathcal{D}_K$ of order zero with supports given by $\cap_{m,n}$, $\cup_{m,n}$. Put $\Phi_{KZ}^{(i)} = \Phi_{KZ}(hX_{i-1}^{(i)}, ht_{i,i+1})$.

**Theorem 7** *The value of the regularized Kontsevich integral on the elementary graphs is:*

$$Z(X_{i,n}^{\pm}) = g_n (\Phi_{KZ}^{(i)})^{-1} \exp(\pm \frac{h}{2} t_{i,i+1}) D_{\sigma_{i,i+1}} \Phi_{KZ}^{(i)} g_n^{-1} \tag{7.14}$$

$$Z(\cap_{i,n}) = g_{n-2} \left( \cap'_{i,n} \Phi_{KZ}^{(i)} \right) g_n^{-1} \tag{7.15}$$

$$Z(\cup_{i,n}) = g_n \left( (\Phi_{KZ}^{(i)})^{-1} \cup'_{i,n} \right) g_{n-2}^{-1}. \tag{7.16}$$

Comparing with the results obtained with the functor $F_U : \mathcal{R} \to \mathcal{D}_K[[h]]$ in theorem 1, we see that for closed ribbon graphs $L$ (framed links), $\hat{Z}(L) = F_U(L)$. This gives another proof that $\hat{Z}$ is an invariant of ribbon graphs.

*Proof.* Let $(z_1, \ldots, z_n) \in \Delta_n$ and consider a path in $V_n$:

$$X_{i,n}^{\pm}(t) = (z_1(t), \ldots, z_n(t)), \tag{7.17}$$

associated to the ribbon graph $X^{\pm}_{i,n}$. We choose the parametrization $z_j(t) = z_j$ if $j \neq i, i+1$ and

$$z_i(t) + z_{i+1}(t) = z_i + z_{i+1}, \tag{7.18}$$

$$z_i(t) - z_{i+1}(t) = e^{\pm i\pi t}(z_i - z_{i+1}). \tag{7.19}$$

In order to compute the corresponding Konsevitch integral, we move away to infinity all strands surrounding the crossing. More precisely, we move first the $n$-th strand on the right, then the $n-1$-th, and so on until the strand $i+2$. Then we are left with the diagram $X^{\pm}_{i,i+1}$. Therefore we write $X^{\pm}_{i,n} = X^{\pm}_{i,n-1} \square \mathrm{id}_1 = (X^{\pm}_{i,n-2} \square \mathrm{id}_1) \square \mathrm{id}_1 = \cdots = X^{\pm}_{i,i+1} \square_{\mathrm{st}} \mathrm{id}_{n-i-1}$, and apply the tensorization theorem (6.22) $n-i-1$ times to get:

$$Z(X^{\pm}_{i,n}) = (D^{(n)}_{n-1} \cdots D^{(i+2)}_{i+1}) Z(X^{\pm}_{i,i+1}) (D^{(n)}_{n-1} \cdots D^{(i+2)}_{i+1})^{-1}. \tag{7.20}$$

After that we send to infinity the distance between the strands $i-1$ and $i$, which translates into further conjugation of $Z(X^{\pm}_{i,i+1})$ by $D^{(i+1)}_{i-1}$. We then use the value of the Kontsevich integral on an isolated crossing: $Z(X^{\pm}_{1,2}) = \exp(\pm(h/2)t_{12}) D_{\sigma_{1,2}}$, and the fact that $(z_{i+1} - z_i)^{\hbar t_{i,i+1}} D^{(i-1)}_{i-2} \cdots D^{(2)}_1$ commutes with $t_{i,i+1}$ to obtain:

$$Z(X^{\pm}_{i,i+1}) = \tilde{g}_{i+1} \exp(\pm \frac{h}{2} t_{i,i+1}) D_{\sigma_{i,i+1}} \tilde{g}^{-1}_{i+1}. \tag{7.21}$$

Now (7.14) follows from theorem 6.

Next we prove (7.15). Proceeding as in the case of crossings, we successively move away from the maximum the strands $n, n-1, \ldots, i+2$, then we send to infinity the distance between the strands $i-1$ and $i$ on the bottom of the maximum, noting that this last operation is not required on the top. We apply the tensorization theorem to $\bigcap_{i,n} = \mathrm{id}_{i-1} \square_{\mathrm{st}} \bigcap_{1,2} \square_{\mathrm{st}} \mathrm{id}_{n-i-1}$, and we get:

$$Z(\bigcap_{i,n}) = (D^{(n-2)}_{n-3} \cdots D^{(i)}_{i-1}) Z(\bigcap_{1,2}) (D^{(n)}_{n-1} \cdots D^{(i+2)}_{i+1} D^{(i+1)}_{i-1})^{-1} \tag{7.22}$$

$$= (D^{(n-2)}_{n-3} \cdots D^{(i)}_{i-1} D^{(i-1)}_{i-2} \cdots D^{(2)}_1) Z(\bigcap_{1,2}) (D^{(n)}_{n-1} \cdots D^{(i+2)}_{i+1} D^{(i+1)}_{i-1} D^{(i-1)}_{i-2} \cdots D^{(2)}_1)^{-1}.$$

Since the value of the Kontsevich integral on an isolated maximum is $Z(\bigcap_{1,2}) = \bigcap'_{1,2} (z_2 - z_1)^{\hbar t_{12}}$, we arrive at (7.15). The proof of (7.16) is similar. $\square$

**Acknowledgements.** We would like to thank P. Degiovanni and J.-M. Maillet for many useful discussions. D. A. also thanks ENSLAPP for its hospitality.